\documentclass[aps,physrev,twocolumn,groupedaddress,showkeys]{revtex4-2}

\usepackage{amsfonts,amsmath,graphicx,natbib,url,hyperref,nicefrac}
\usepackage{amssymb, verbatim, paralist, soul, comment}
\usepackage{ tabularx, booktabs}

\usepackage{float}

\usepackage{tikz}
\usepackage{color}

\usepackage{subfig}

\usepackage[T1]{fontenc}
\usepackage{aecompl}
\usepackage{calligra}
\DeclareMathAlphabet{\mathcalligra}{T1}{calligra}{m}{n}
\DeclareFontShape{T1}{calligra}{m}{n}{<->s*[2.2]callig15}{}

\usepackage{hyperref}
\hypersetup{
    pdfnewwindow=true,      
    colorlinks=true,       
    linkcolor=orange,          
    citecolor=cyan,        
    filecolor=blue,      
    urlcolor=blue           
}

\DeclareMathAlphabet{\mathcalligra}{T1}{calligra}{m}{n}
\DeclareFontShape{T1}{calligra}{m}{n}{<->s*[2.2]callig15}{}

\usepackage{color}

\usepackage{comment}

\begin{document}

\title{Instability and Angular Momentum Transfer in Thick Circumbinary Disks}

\author{Allen R. Murray}
\affiliation{Department of Physics and Astronomy, Purdue University, 525 Northwestern Avenue, West Lafayette, IN 47907, USA}
\email{murray92@purdue.edu}

\author{Paul C. Duffell}
\affiliation{Department of Physics and Astronomy, Purdue University, 525 Northwestern Avenue, West Lafayette, IN 47907, USA}
\email{pduffell@purdue.edu}

\date{\today}

\begin{abstract}
Utilizing the \texttt{DISCO} code we solve the 2D hydrodynamic equations for an equal mass circular restricted binary with an isothermal circumbinary disk (CBD). This work explores the eccentric instability in the CBD and angular momentum transfer, and their dependence on disk thickness and viscosity.  The focus is specifically on thicker CBDs, with Mach number, ($\mathcal{M}\leq 14$). The growth rate of the disk cavity's eccentricity and the angular momentum transport from accretion is calculated as diagnostics for binary-disk morphology evolution. Calculations of CBD torques must run long enough to allow saturation of the disk's eccentric instability.  The torque on the binary rapidly flips sign once the instability saturates. Thus we connect eccentricity saturation to a reversal in migration behavior for the binary. The binary experiences inward migration while the disk is in an evolving state, and outward migration once the disk eccentricity has saturated. The saturation time was found to be non-linearly dependent on Mach number and thicker CBDs require longer evolution time for a quasi-steady state to appear.  The eccentric instability is not present for $\mathcal{M} \lesssim 5$, and therefore migration reverses direction, leading to robustly inward migration for very thick disks $h/r \gtrsim 0.2$. The results of this work were found to be consistent with community results for Mach numbers $\mathcal{M} \ge 10$.
\end{abstract}


\maketitle

\section{Introduction}

Circumbinary Disks (CBD) are a natural phenomenon occurring in many astrophysical contexts. For example, during the chaotic and violent process of galaxy mergers, the Super Massive Black Hole Binary (SMBHB) potentially formed should also form a CBD of gas that is thought to drive accretion onto the binary (\cite{Begelman1980}; \cite{Milo2001}; \cite{Cuadra2009}; \cite{Chapon2013})) and cause orbital inspiral at distances larger than required for efficient gravitational wave emission \cite{Haiman2009}. Additionally, CBDs can form through fallback accretion around a massive giant star and white dwarf companion (e.g., \cite{Kashi2011}; \cite{Tuna2023}), and through disk fragmentation in young binary star evolution (e.g., \cite{Boss1986}; \cite{Bonnell1994a}; \cite{Krater2008}). \\
Extensive numerical work has been done to study CBD systems across a vast range of binary-disk parameters. One striking morphological feature seen in a wide range of computational studies is a low density eccentric cavity around the binary (\cite{MacFad2008}, \cite{Munoz2019}; \cite{Duffel2020}; among others). This eccentric cavity is present for binary mass ratio exceeding 0.04, (see \cite{DOrazio2016} for discussion on cavity size dependence on mass ratio) and is often accompanied by an over-density at the cavity edge, colloquially called the "lump" or "clump". This clump is thought to be a by-product of the growth of cavity eccentricity (\cite{ShiKrolikLubow2012}, called SKL12 hereafter), and can cause a spike in accretion at a frequency close to the orbital frequency at the cavity edge  (SKL12; \cite{Dorazio2013}; \cite{Farris2014}; \cite{Duffel2020}).  \\
Some of the earliest predictions \cite{ArtyLubow94} found that interaction with the CBD, through either accretion or resonances, caused the binary to lose angular momentum and migrate closer.  This was supported by most numerical studies, until it was noticed by \cite{Tang2017} that outward migration occurred in some models.  In that study it was assumed that this was a numerical artifact, but later \cite{Moody2019} and \cite{Munoz2020} consistently found outward migration, measured both via binary torques and from the inward flow of angular momentum in the disk toward the binary.  This was later supported by complimentary findings by \cite{Duffel2020}, and it was suggested that outward migration might be the generic outcome of binary accretion.  For this range of parameters, the topic of outward migration has been reasonably settled by the code comparison of \cite{SantaBarbara24}, in which eleven different numerical schemes all found outward migration for a disk with $h/r = 0.1$.\\
However, this might not be the generic outcome of binary accretion.  \cite{Tiede2020} (Hereafter T20) first showed a transition between outward and inward migration for sufficiently thin disks ($h/r \lesssim 0.04$). \cite{HeathNixon2020} found a similar result for 3-D disks. \cite{DittmannRyan22} (Hereafter, DR22) found a similar transition in migration behavior for very thin disks.  Thus, the migration of an accreting binary may be quite sensitive to the Mach number of the disk.  \\
So far, most studies have concentrated on disks that were at least moderately thin.  In other words, the thick disk regime $h/r > 0.1$ has not been explored systematically.  This work will investigate the behavior of the CBD system with an equal mass circular binary in the thick disk regime. Specifically, the varied disk parameters are Mach number ($\mathcal{M} \leq 14$, $h/r \geq 0.07$) and viscosity. For discussion of binary orbital evolution in the eccentric equal mass case see \cite{Zrake2021}, and \cite{DOrazio2021}. \\
This paper is structured as follows: \autoref{sec:num} will detail our methods and numerical setup, \autoref{sec:results} contains our results, and \autoref{sec:sum} will state our conclusions.

\section{Numerical Methods}
\label{sec:num}

The $\texttt{DISCO}$ hydrodynamical modeling code in the 2-D($r$,$\phi$) vertically integrated configuration \citep{DISCO} was utilized to solve the viscous hydrodynamic equations for a gaseous disk orbiting an equal mass circular restricted binary. $\texttt{DISCO}$ uses a moving mesh scheme to model the disk accurately over an extended time period. The disk is set to be isothermal with a constant kinematic viscosity prescription.  We use code units such that $G = M_B = a_B = 1$ in order for $\Omega_B$ to also be unity and the time is measured in terms of binary orbits times a factor of 2$\pi$. 

\begin{figure}
    \includegraphics[width=0.50\textwidth]{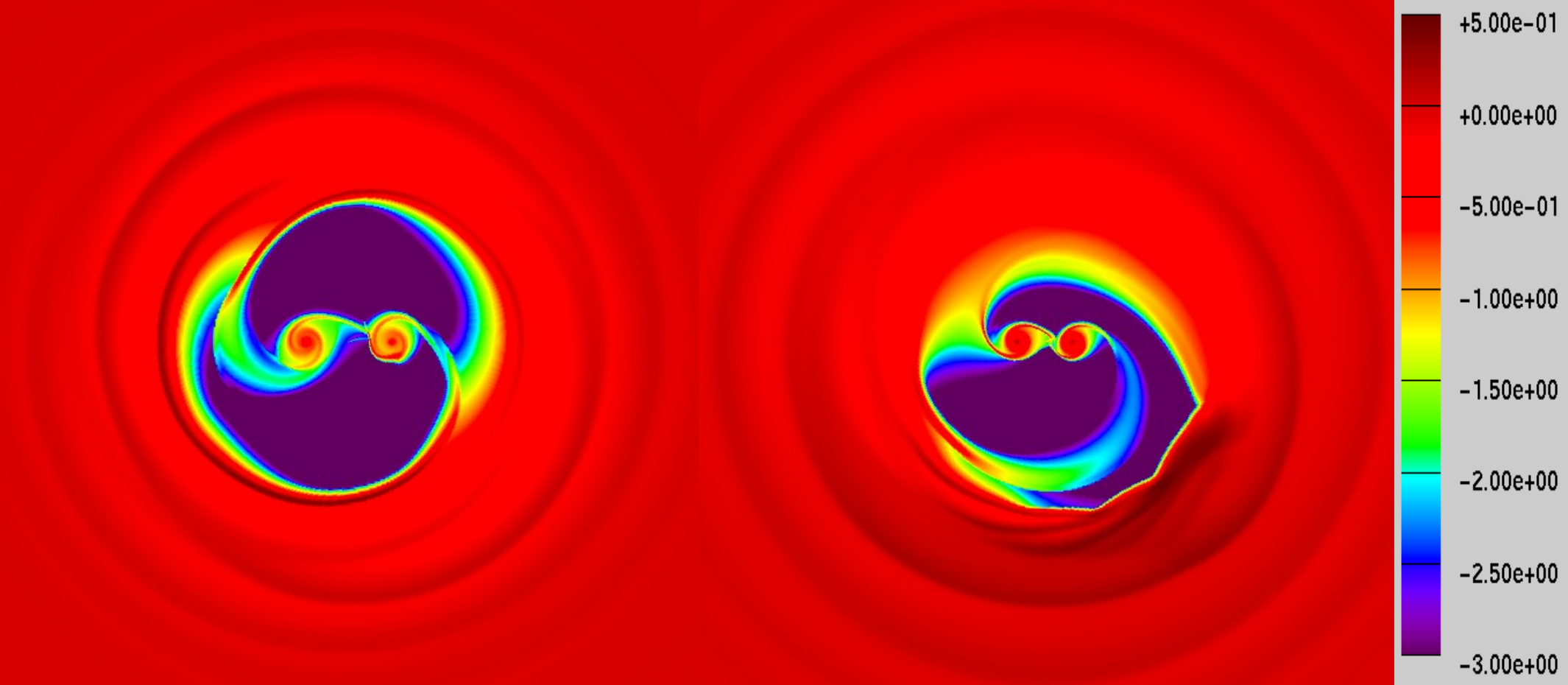}
    \caption{Snapshots from \texttt{DISCO} for Mach number, $\mathcal{M} = 11.25$ with the fiducial viscosity $\nu = 10^{-3}a_B^2\Omega_B$. Shown is the surface density of gas in the disk. The saturated state of the eccentric instability at $t = 1000$ orbits (right) is contrasted with the still evolving state at $t = 25$ orbits (left). The saturated state also is marked by the presence of "the clump". }
    \label{fig:fig1}
\end{figure} 

These results were standardized to the specifications outlined in the numerical code review \citep{SantaBarbara24} and checked for agreement with their results. The specifications are repeated below for ease of reference.
The following equation for the initial surface density profile of the disk was utilized,
\begin{equation}
    \Sigma(r) = \Sigma_0[(1-\delta_0)e^{-(R_{cav}/r)^{12}}+\delta_0]f(r)
\end{equation}
where $R_{cav}=2.5a_B$, the radius of the cavity, and $\delta_0 = 10^{-7}$, the initial density for the cavity. f(r) is a truncation function for a smooth outer boundary and given by the following,
\begin{equation}
    f(r) = 1 - \frac{1}{1+e^{-2(r-R_{out}/a_B)}}
\end{equation}
where $R_{out}=10a_B$. A locally isothermal equation of state is employed with sound speed,
\begin{equation}
    c_s^2 = -\Phi_B(t;r,\phi)/\mathcal{M}^2.
\end{equation}
Where $\Phi_b$ is the gravitational potential of the binary with a softening parameter $\epsilon$,
\begin{equation}
    \Phi_B = \sum_j \Phi_j = \sum_j \frac{-GM_j}{\sqrt{|r_{ij}|^2 +\epsilon^2}}
\end{equation}
and $\mathcal{M}$ is the azimuthal Mach number given by,
\begin{equation}
    \mathcal{M} = \frac{r\Omega(r)}{c_s} = \frac{r}{H}.
\end{equation}

\begin{figure}
    \centering
    \includegraphics[width=0.50\textwidth]{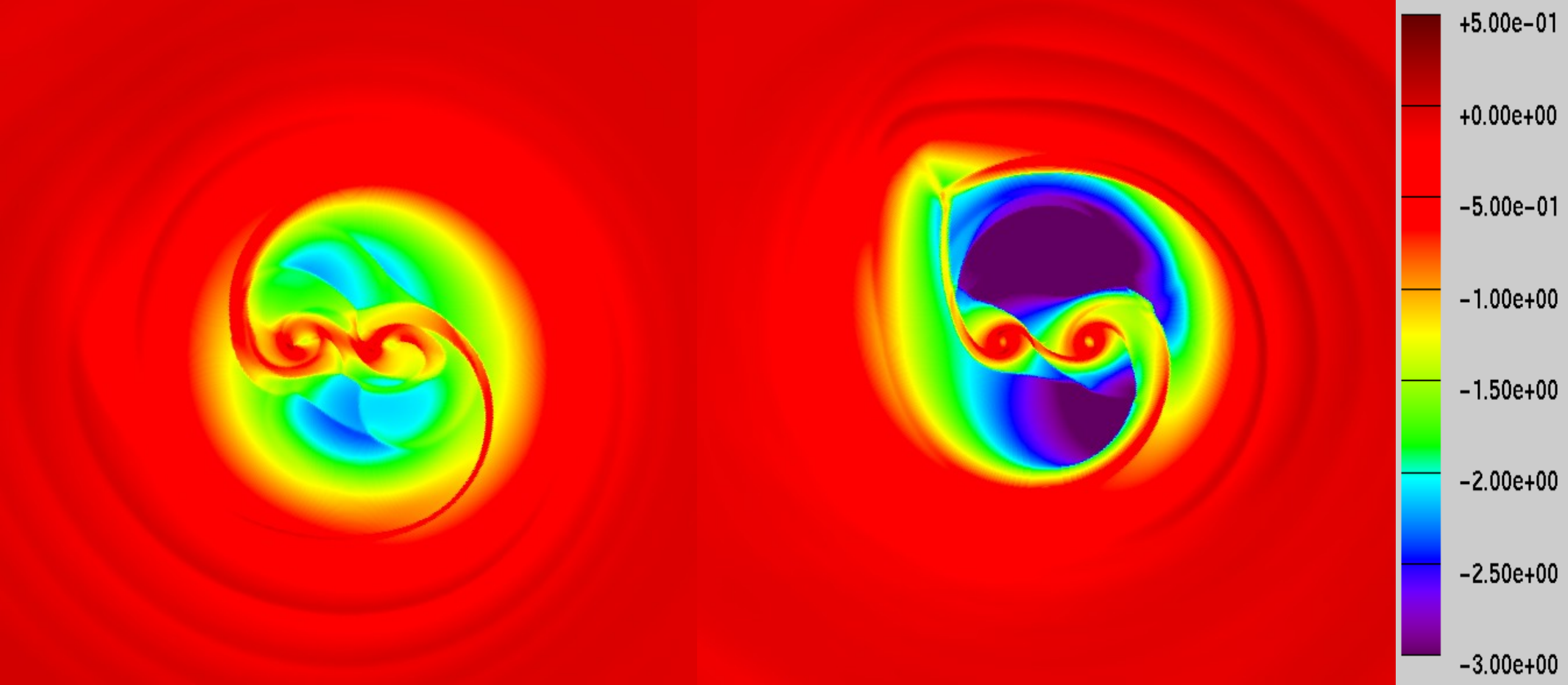}
    \caption{Snapshot from DISCO for Mach number $\mathcal{M} = 4$ (left) and $\mathcal{M} = 5.5$ (right) with fiducial viscosity at $t = 5000$ orbits. Note that the disk in the left panel has a shallower cavity than the right and the cavity morphology remains circular.}
    \label{fig:fig 2}
\end{figure}

The initial angular frequency of gas in the disk is set to the equilibrium solution far from the binary,
\begin{equation}
    \Omega_0^2(r) = \frac{GM_B}{r}\left( 1-\frac{1}{\mathcal{M}^2}\right),
\end{equation}
and is modified such that it flattens near the binary,
\begin{equation}
    \Omega(r) = [\Omega_0(r)^{-4} + \Omega_B^{-4}]^{-\frac{1}{4}}.
\end{equation}
In this study, a uniform kinematic viscosity $\nu$ is utilized with the fiducial value being,
\begin{equation}
    \nu = 10^{-3}a_B^2\Omega_B.
\end{equation}
An initial radial velocity kick to the cavity is introduced of the form,
\begin{equation}
    v_r(r) = v_0 {\rm sin}(\phi)(\frac{r}{a_B}) {\rm exp}\left[-\left(\frac{r}{3.5a_B}\right)^6\right],
\end{equation}
with $v_0 = 10^{-4}\Omega_Ba_B$ in order to seed the eccentric elongated cavity. As stated in \cite{SantaBarbara24}, the initial seed to the cavity is not needed for eccentricity growth, as numerical noise would be sufficient, but as the focus of this work is to study the growth of this instability explicitly, convergence of the solution will necessitate a non-zero seed perturbation.

The system is initialized with 720 radial zones, split with linear scaling from the origin until $R = a_B$, and logarithmic scaling at large radii. The disk has an inner radius $R_{in} = 1a_B$ and outer edge at $R_{out} = 30a_B$. The azimuthal zones were varied per radial slice in order to maintain a roughly uniform aspect ratio. This corresponds to a resolution of $\delta r = 0.02a_B$ at $r = 3a_B$. 

\begin{figure*}
    \centering
    \includegraphics[width=0.9\textwidth]{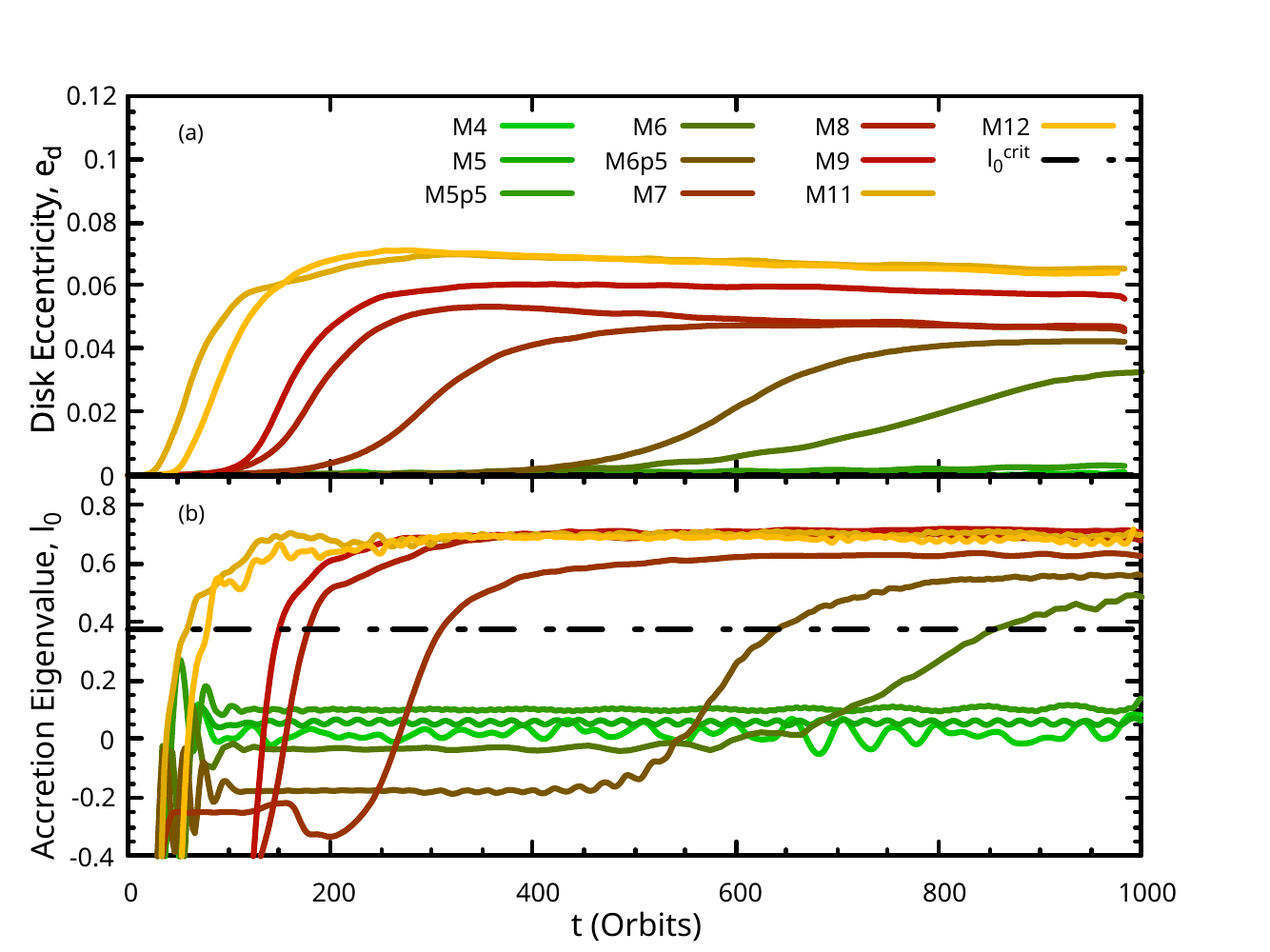}
    \caption{Time evolution of the disk eccentricity (top) and accretion eigenvalue, $l_0$ (bottom) for 1000 binary orbits. Each curve corresponds to a different Mach number. The evolution of accretion eigenvalue shows that initially the binary experiences inward migration but once the disk eccentricity saturates the binary switches to outward migration.}
    \label{fig:fig 3}
\end{figure*}

In this study we classify two states of the binary-disk system. One, where the cavity eccentricity grows exponentially over time, called the evolving state, and two, where the cavity eccentricity has reached its peak value, called the saturated state. \autoref{fig:fig1} shows surface density snapshots from \texttt{DISCO} for our fiducial viscosity $\nu = 10^{-3}a_B^2\Omega_B$ at separate times with Mach number $\mathcal{M} = 11.25$. The left panel shows the disk while it is in the evolving state, and the right panel is the saturated eccentric state of the disk.

The growth rate of the cavity's eccentricity is characterized using $\Gamma$, defined as,
\begin{equation}
    \Gamma = \frac{|\dot{e}_d|}{|e_d|},
\end{equation}
where $e_d$ is calculated with the same method as \cite{SantaBarbara24}. The reported growth rates were calculated using the average of the gaussian-convolved growth rate over the observed period of exponential growth. These growth rates are reported in \autoref{tab:1}. 
We also seek to connect the growth of the eccentric cavity to the orbital behavior of the binary. The semi-major axis evolution equation from \cite{MirMunLai2017} was adopted, and reprinted below as,
\begin{equation}
    \frac{{
{\dot a}_B}}{a_B} = 8 \left(\frac{l_0}{l_B} - \frac{3}{8} \right)\frac{\dot{M}}{M_B}.
\end{equation}
Where $\dot{M}/M_B$ is the (normalized) accretion rate and $l_0/l_B$ is the accretion eigenvalue in units of the characteristic specific angular momentum ($l_B = a_B^2\Omega_B$). Thus, the binary is expected to migrate outwards if $l_0/l_B > 3/8$ and migrate inwards when $l_0/l_B < 3/8$. Therefore, $l_0^{\rm crit}=0.375$, and the accretion eigenvalue is calculated for an equal mass circular binary as,
\begin{equation}
    \frac{l_0}{l_B} = \frac{\bar{T}_{grav} + \dot{J}_{acc}}{\dot{\bar{M}}}.
    \label{eq:eq12}
\end{equation}
Where, $\bar{T}_{grav}$ is the gaussian convolved gravitational torque on the binary, $\dot{\bar{M}}$ the accretion rate, and $\dot{J}_{acc}$ is the angular momentum of material removed by the sinks. In the small sink limit, $\dot{J}_{acc} = 0.25\dot{\bar{M}}$ for a circular equal mass binary and our accretion eigenvalue equation becomes:

\begin{figure*}[ht]
    \includegraphics[width=0.9\textwidth]{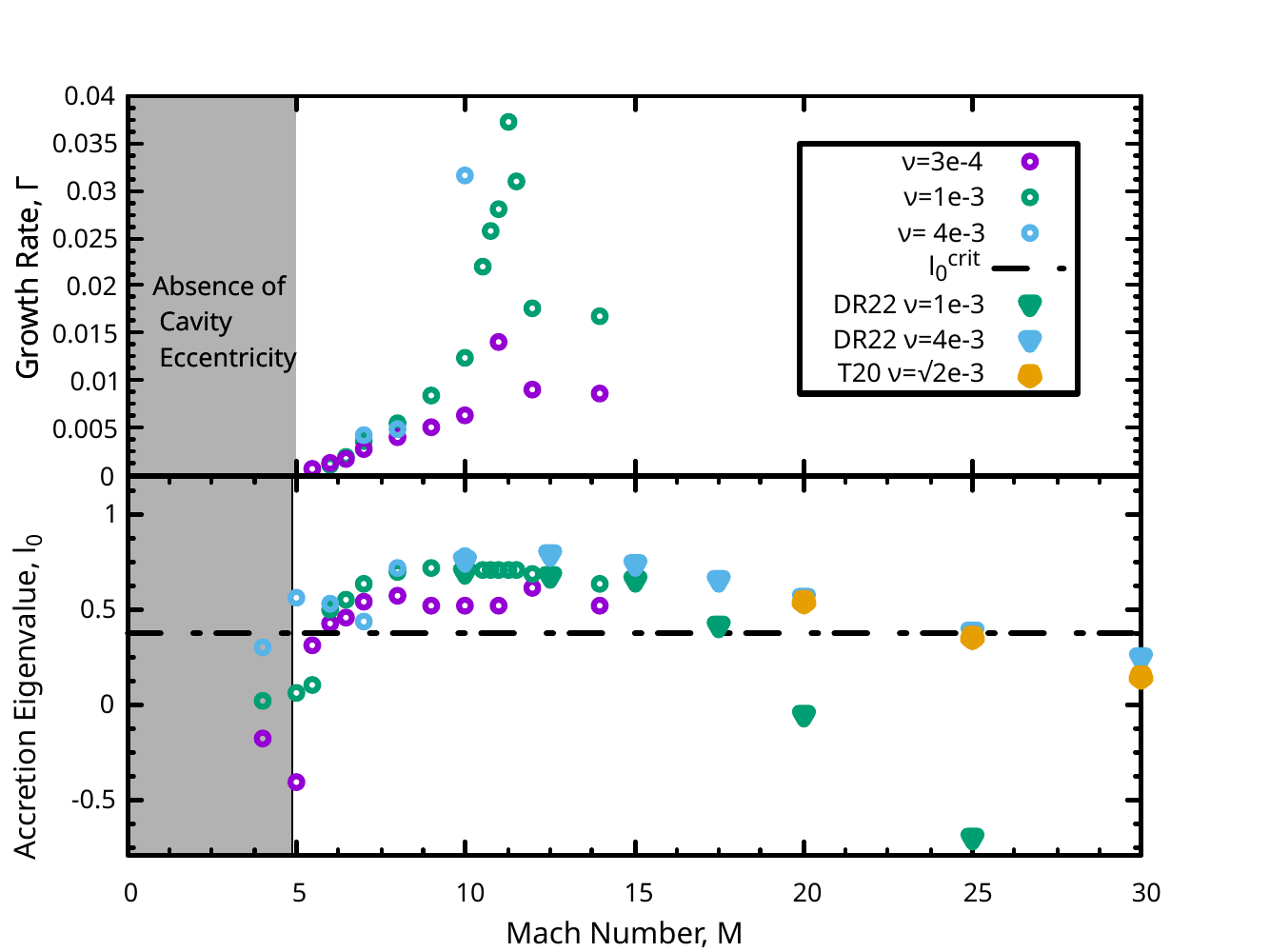}
    \caption{Top: Growth rate of the disk's eccentricity as a function of Mach number, and disk viscosity.  A peak in growth rate was found around $\mathcal{M} \approx 11.25$ and as $\mathcal{M}$, and a stable regime with zero growth for $\mathcal{M} < 5$. Bottom: Late-time accretion eigenvalue as a function of Mach number and disk viscosity. This is compared with $l_{0_{crit}} = 0.375\sqrt{GMa_B}$, to determine whether binary migration is inward or outward. $l_0$ attains a maximum near the community result ($l_0 \approx 0.7\sqrt{GMa_B}$ for fiducial viscosity. Plotted as well is $\frac{d\rm ln(a_B)}{d\rm ln(M)}$ from DR22 for various viscosities and accretion eigenvalue from T20 with their respective Mach numbers. Good agreement is found in the overlap between the respective parameter spaces.}
    \label{fig:fig 4}
\end{figure*}

\begin{equation}
    \frac{l_0}{l_B} = \frac{\bar{T}_{grav}}{\dot{\bar{M}}} + 0.25
\end{equation}

This work will determine the growth rate of the cavity's eccentricity and the accretion eigenvalue as the Mach number and kinematic viscosity are varied. These quantities were measured while evolving the system for a minimum 1000 orbits as to allow the disk to reach the saturated state. In order to properly account for long period effects the system was evolved for several viscous times. The low viscosity runs $(\nu = 3\times 10^{-4}a_B^2\Omega_B)$ required a minimum of 2000 orbits for the saturated state to be reached. In addition, some low Mach number, fiducial viscosity runs $(\mathcal{M} \leq 6)$ were run for 5000+ orbits in order to obtain a saturated $l_0$ value. Below $\mathcal{M} = 5$ even at 5000-10,000 orbits the disk remains circular, as shown in \autoref{fig:fig 2}. Where the surface density of the fiducial viscosity, $\mathcal{M} = 4$ model is shown in the left panel and $\mathcal{M} = 5.5$ in the right panel at $t = 5000$ orbits. The cavity on the left remains circular and is notably shallower than the cavities seen in \autoref{fig:fig1} and the right panel. 

\section{Results}
\label{sec:results}


\autoref{fig:fig 3} presents the evolution of disk eccentricity $e_d$ and accretion eigenvalue $l_0$ over time.  First, note that in all models, $l_0$ is initially below $3/8$, giving inward migration.  Then, after some time the value of $l_0$ suddenly flips to an outwardly-migrating value, with $l_0 > 3/8$. This work found a clear connection between the saturation of the eccentric instability and the flip in migration behavior for the binary.  This can be seen most clearly by noting the time of eccentricity saturation is coincident with the time that $l_0$ flips from inward to outward in all cases.

\begin{table*}[ht]
    \begin{tabular}{| c | c c c c | c | c c c c | c | c c c c |}
    \hline
       Run Name & $\nu $  & $\mathcal{M}$ & $\Gamma$ & $l_0$ & Run Name & $\nu$ & $\mathcal{M}$ & $\Gamma$ & $l_0$ & Run Name & $\nu$ & $\mathcal{M}$ & $\Gamma$ & $l_0$ \\
    \hline
    LVM4 & 3e-4 & 4 & 0 & 0 & FVM4 & 1e-3 & 4 & 0 & 0.0204 & HVM4 & 4e-3 & 4 & 0 & 0.2960 \\ 
    \hline
    LVM5 & 3e-4 & 5 & 0.00061 & -0.4113 & FVM5 & 1e-3 & 5 & 0 & 0.0577 & HVM5 & 4e-3 & 5 & 0 & 0.5545 \\
    \hline 
    LVM5p5 & 3e-4  & 5.5 & 0.00064 & 0.3092 & FVM5p5 & 1e-3 & 5.5 & 0 & 0.1024 &  &  &  &  &  \\
    \hline
    LVM6 & 3e-4  & 6 & 0.00138 & 0.3091 & FVM6 & 1e-3 & 6 & 0.00122 & 0.5014 & HVM6 & 4e-3 & 6 & 0 & 0.5238 \\
    \hline 
    LVM6p5 & 3e-4 & 6.5 & 0.00181 & 0.4587 & FVM6p5 & 1e-3 & 6.5 & 0.00196 & 0.5467 &  &  &  &  &  \\
    \hline
    LVM7 & 3e-4 & 7 & 0.00208 & 0.5336 & FVM7 & 1e-3 & 7 & 0.00358 & 0.6287 & HVM7 & 4e-3 & 7 & 0.00416 & 0.4320 \\
    \hline 
    LVM8 & 3e-4 & 8 & 0.00407 & 0.5664 & FVM8 & 1e-3 & 8 & 0.00559 & 0.6939 & HVM8 & 4e-3 & 8 & 0.00495 & 0.7163 \\
    \hline
    LVM9 & 3e-4 & 9 & 0.00519 & 0.5134 & FVM9 & 1e-3 & 9 & 0.00833 & 0.7125 &  &  &  &  &  \\
    \hline 
    LVM10 & 3e-4 & 10 & 0.00630 & 0.5172 & FVM10 & 1e-3 & 10 & 0.0124 & 0.7164 & HVM10 & 4e-3 & 10 & 0.03166 & 0.7753 \\
    \hline
     &  &  &  &  & FVM10p5 & 1e-3 & 10.5 & 0.0219 & 0.7006 &  &  &  &  &  \\
    \hline
     &  &  &  &  & FVM10p75 & 1e-3 & 10.75 & 0.0258 & 0.7053 &  &  &  &  &  \\
    \hline
    LVM11 & 3e-4 & 11 & 0.011404 & 0.5205 & FVM11 & 1e-3 & 11 & 0.0281 & 0.7012 &  &  &  &  &  \\
    \hline
     &  &  &  &  & FVM11p25 & 1e-3 & 11.25 & 0.0372 & 0.7044 &  &  &  &  &  \\
    \hline
     &  &  &  &  & FVM11p5 & 1e-3 & 11.5 & 0.00309 & 0.7042 &  &  &  &  &  \\
    \hline
    LVM12 & 3e-4 & 12 & 0.00908 & 0.6119 & FVM12 & 1e-3 & 12 & 0.0176 & 0.6858 &  &  &  &  &  \\
    \hline
    LVM14 & 3e-4 & 14 & 0.00867 & 0.5229 & FVM14 & 1e-3 & 14 & 0.0168 & 0.6368 &  &  &  &  &  \\
    \hline
    \end{tabular}
    \caption{Complete simulation suite that is employed in this work. Units for the listed quantities are, $\nu$ [$a_B^2\Omega_B$], $\Gamma$ [$\Omega_B^{-1}$], and $l_0$ [$\sqrt{GMa_B}$]. }
    \label{tab:1}
\end{table*}

Second, it was found that with decreasing Mach number the circumbinary disk takes an increasingly longer time to reach the saturated state. 
When Mach number is about 5 ($\mathcal{M}\lesssim 5$ or $h/r \gtrsim 0.2$), the growth rate is negligible and the cavity remains symmetric. In order to confirm that the instability is no longer present, models FVM4, FVM4p5, and FVM5 were run for 5000+ orbits and these systems had no appreciable growth in the eccentric instability. Notably, the cavity is shallower in this regime which may point towards a transition to a pressure dominated disk as suggested by \cite{DOrazio2016}, section 2.3.1.  

Our full results are listed in \autoref{tab:1}, along with our naming scheme and model parameters. The blank spaces represent gaps in our model suite.  Across the model suite it's found that this lower critical Mach number is approximately 5 ($h/r \approx 0.2$). This is plotted as the gray region in \autoref{fig:fig 4} along with out measured eccentric instability growth rates (top) and accretion eigenvalue, $l_0$ (bottom) as functions of disk Mach number and disk viscosity. Larger viscosity is found to lead to slightly faster growth, but the growth rate is only weakly dependent on viscosity in this regime. \\



Comparing the results of this work with similar systematic studies yields good agreement with results from T20 and DR22. This comparison is plotted in \autoref{fig:fig 4}. Colors delineate viscosity and the points from this study are unfilled circles. Accretion eigenvalue approaches the community fiducial value of $l_0 \approx 0.7\sqrt{GMa_B}$ (\cite{MunLaiKrat20};\cite{Duffel2020}) as Mach number ranges from ($7 \leq \mathcal{M} \leq 14$). DR22 used a similar diagnostic for their intermediate Mach number models. For comparison, their results for $\frac{dloga_B}{dlogM}$ were converted into corresponding $l_0$ values using equation 25 from DR22. The points from DR22 are plotted with filled triangles and only the points that have similar disk viscosities to this work were utilized. Also plotted are the higher Mach number results from T20. There is good agreement between these results and DR22 where the two systematic studies overlap. These three studies together provide a clear picture of binary migration transition points as a function of Mach number for $4 \leq \mathcal{M} \leq 30$. 

\section{Conclusions}
\label{sec:sum}

The $\texttt{DISCO}$ code was utilized to solve the 2-D($r$,$\phi$) viscous hydrodynamic equations for an equal mass circular restricted binary with an extended gaseous circumbinary disk for a minimum of 1000 binary orbits. Mach number $\mathcal{M}$, and disk viscosity $\nu$ were varied across the models to determine the growth rate of the eccentric instability and angular momentum transportation's dependence on these two parameters. This study focuses specifically on the low mach number or thick disk regime. \\
Listed below are the main results across the model suite of this work.
\begin{enumerate}
    \item First, when Mach number is below 5, ($\mathcal{M}\lesssim 5$, or $h/r \gtrsim 0.2$) there is an absence of eccentricity and a notably shallower cavity.
    \item Second, a connection between the saturation of the eccentric instability in the disk and a change in migration behavior of the binary is reported. The binary experiences outward migration when the disk is in the saturated eccentric state, and inward migration when the cavity is still circular.
    \item Third, it's found that growth rate and accretion eigenvalue are at most weakly dependent on disk viscosity in this regime. 
    \item Finally, the results of this work agree with other studies in the region of overlap of our parameter space. Specifically, T20 and DR22.
\end{enumerate} 

\begin{acknowledgements}
    We are grateful to C. Tiede, D. D'Orazio, and J. Zrake for their insightful suggestions and discussion. This research was supported by the National Science Foundation under grant number AAG-2206299.
\end{acknowledgements}
\bibliography{main.bib}

\end{document}